\documentclass[twocolumn,nofootinbib,linenumbers]{revtex4}

\usepackage{graphicx}  
\usepackage{lipsum}
\usepackage{siunitx}
\usepackage{booktabs} 
\usepackage{float}

\usepackage[T1]{fontenc}
\usepackage{multirow}
\usepackage{array}
\usepackage{caption}
\usepackage{graphicx}

\usepackage{bm}        
\usepackage{amsfonts}  
\usepackage{amsmath}   
\usepackage{amssymb}   
\usepackage{subcaption}
\usepackage{lineno}
\usepackage{hyperref}
\hypersetup{colorlinks=true}

\begin{document}

\title{Searching for HWW Anomalous Couplings with Simulation-Based Inference}

\author{Marta Silva}
\email[\href{mailto:femarta@cern.ch}{femarta@cern.ch}]{}
\affiliation{Nikhef National Institute for Subatomic Physics, Amsterdam, The Netherlands}
\affiliation{Laboratory of Instrumentation and Experimental Particle Physics, Lisbon, Portugal}

\author{Ricardo Barrué}
\email[\href{mailto:ricardo.barrue@cern.ch}{ricardo.barrue@cern.ch}]{}
\affiliation{Marietta Blau Institute for Particle Physics, Vienna, Austria}
\affiliation{Instituto Superior Técnico, Lisbon, Portugal}
\affiliation{Laboratory of Instrumentation and Experimental Particle Physics, Lisbon, Portugal}

\author{Inês Ochoa}
\email[\href{mailto:miochoa@lip.pt}{miochoa@lip.pt}]{}
\affiliation{Laboratory of Instrumentation and Experimental Particle Physics, Lisbon, Portugal}

\author{Patricia Conde Muíño}
\affiliation{Instituto Superior Técnico, Lisbon, Portugal}
\affiliation{Laboratory of Instrumentation and Experimental Particle Physics, Lisbon, Portugal}

\begin{abstract}

    Understanding the source of the universe’s asymmetry between matter and antimatter is one of the major open questions in particle physics. In this work, the sensitivity of novel machine-learning-based inference techniques to CP-odd and CP-even $HWW$ anomalous couplings is studied in the $WH \rightarrow \ell \nu b\bar{b}$ channel ($\ell = e, \mu$), within the Standard Model Effective Field Theory (SMEFT) framework.

    Two machine-learning simulation-based inference (SBI) methods are explored: a per-event likelihood-ratio estimator, which directly approximates the ratio of probability densities between competing hypotheses, is benchmarked against a per-event optimal-observable estimator optimized for sensitivity to the parameters of interest. Both approaches are also compared to traditional summary statistics, in this case histograms of kinematic and angular observables, as commonly used in experimental analyses.
    
    SBI methods provide tighter constraints than one-dimensional summary statistics, though their performance is comparable to two-dimensional histogram analysis. The optimal-observable approach remains promising for its ability to probe multiple couplings simultaneously. Restricting the analysis to a region of high $S/B$ also enhances sensitivity to CP-odd operators while preserving sensitivity to CP-even operators, which histogram analyses often lose. Although the likelihood-ratio estimator sometimes struggles with likelihood minima and shapes, optimisations that target its robustness could make it more sensitive than both the optimal-observable estimator and the histogram method. These results underscore the potential of advanced simulation-based inference techniques, encouraging further exploration with LHC Run 3 data to surpass current ATLAS and CMS sensitivities.

\end{abstract}

\maketitle

\section{Introduction}

One of the major questions left unaddressed by the Standard Model (SM) is the observed asymmetry between matter and antimatter in the Universe. Sources of charge-parity (CP) symmetry violation in the SM are insufficient to explain this imbalance~\cite{Andrei_D_Sakharov_1991} and CP violation in the Higgs sector is a well-motivated way to address it. In fact, since the discovery of the Higgs boson, one of the main goals of the Large Hadron Collider (LHC) has been to precisely measure its properties.

Most analyses in experimental particle physics share a common central question: \textit{based on observed data, what is the most likely value of the physical parameters that describes the underlying process?}. This is known as statistical inference and its fundamental key is the likelihood function. However, in particle physics this quantity is typically intractable as it involves integrating over millions of random variables~\cite{article_SBI_nature}. Performing statistical inference when the data-generating process does not have a tractable likelihood is a major challenge common to many fields, and the methods that address this intractability using simulators of the underlying physica process are collectively referred to as simulation-based inference techniques.

The traditional approach to search for beyond the SM (BSM) physics, such as anomalous couplings in the Higgs sector, is to reduce the dimensionality of the probability density function by filling histograms or templates from Monte Carlo (MC) simulations of one or two discriminating observables (summary statistics). Common examples are angular variables which are used e.g. by the ATLAS Collaboration to constrain the presence of CP-violating components in Higgs and EW boson interactions via the $H\to WW^{*}$~\cite{ATLAS-HWW} decay or in $WH$~\cite{ATLAS-WH-PUB} production. However, while this is straightforward and computationally efficient, it leads to an inevitable loss of information about the process and underlying physical parameters being studied \cite{Brehmer_2020, Castro:2022zpq}.

Other widely used methods, such as the Matrix Element Method (MEM) \cite{Kondo:1991dw} or Optimal Observables \cite{Diehl:1993br}, used by ATLAS and CMS e.g. in measurements of vector boson fusion (VBF) production of $H\to 4 \ell$~\cite{ATLAS-H4l,CMS-H4l} and $H\to\tau\tau$~\cite{Htautau-CP,CMS:MELA_Htautau} leverage the fact that some parts of the likelihood function can be computed, particularly through matrix element calculations, to compress the information from multiple kinematic variables with minimal losses in sensitivity. Nevertheless, these methods rely on approximations of the parton shower, hadronization, and detector interactions, thereby simplifying the typical $pp$ collision simulation chain. 

More recently, Machine Learning (ML) methods have been recognized as valuable tools for handling this problem without relying on simplifications or compromising statistical power. In particular, novel ML-inference techniques, introduced in Refs. \cite{PhysRevLett.121.111801,Brehmer:2018eca, Brehmer:2018hga, Mastandrea:2024irf, Bahl:2024meb, Ghosh:2025fma}, exploit additional information from MC simulations in order to train Neural Networks (NN) that can effectively estimate quantities relevant to statistical inference. An example is the likelihood ratio which, according to the Neyman-Pearson lemma, is the optimal test statistic for simple-vs-simple hypothesis testing~\cite{Neyman:1933wgr}. 

In this work, we study the sensitivity of these innovative techniques to $HWW$ anomalous couplings with the main goal of designing the best strategies to employ in the analysis of LHC Run 3 data and beyond, highlighting areas of potential future improvements. We focus in particular on the $WH$ production channel, with the $W$ boson decaying to leptons and the Higgs boson to a pair of $b$-quarks, with anomalous couplings parametrized with the Standard Model effective field theory (SMEFT)~\cite{BRIVIO20191} framework. In particular, we consider both CP-violating and CP-conserving contributions to the Lagrangian, in order to disentangle their contributions and provide a more complete picture of anomalous couplings to this vertex. Kinematic and angular observables commonly used in experimental analyses are benchmarked against three ML-based methods described in the literature, SALLY~\cite{Brehmer:2018eca}, ALICE and ALICES~\cite{Stoye:2018ovl}. Existing work on this topic includes a study of the SALLY method on the same final state, but focusing solely on the CP-violating coupling~\cite{sally_cpv_wh}, and a study of SALLY for targeting only CP-even couplings~\cite{Brehmer:2019gmn}. Additionally, a practical application of SBI techniques has been carried out by the CMS Collaboration in $WH$ and $ZH$ production~\cite{CMS-WH-SBI}, targeting multiple SMEFT operators. Finally, an implementation of ML-based SBI has been performed by the ATLAS Collaboration, using an off-shell Higgs boson couplings measurement as an example~\cite{ATLAS:2025clx}. Ultimately, our work aims to inform analysis strategies in order to extract the best sensitivity to BSM physics from LHC data, using novel but increasingly robust and widespread SBI approaches.

The remainder of this document is structured as follows: Section \ref{sec:sbi} introduces simulation-based inference and the methods employed in this work. Section \ref{sec:analysis} outlines the effective field theory framework and the anomalous couplings studied, as well as the details of the analysis and event generation. Section \ref{sec:results} presents the results of all the scenarios studied and, finally, in Section \ref{sec:conclusion}, the main conclusions of this work are drawn.

\section{Simulation-based Inference}\label{sec:sbi}

\subsection{The likelihood-free inference problem}

The central quantity in any particle physics analysis is the likelihood function $p_{\text{full}}({x}|\theta)$, which quantifies the probability of observing a set of events, each characterized by a vector of observables $x$ as a function of the parameters of interest $\theta$. In this work, $\theta$ will parameterize the effect of the anomalous couplings, which will be introduced in the next Section. The likelihood function is usually expressed as
\begin{equation}
    p_{\text{full}}({x}|\theta) = \si{Pois} (n | L \sigma (\theta)) \prod_{i} p(x_i|\theta),
\end{equation}
where $n$ represents the observed number of events, $L$ the integrated luminosity, $\sigma(\theta)$ the cross section, \mbox{$\text{Pois}(n|\lambda) = \lambda^n e^{ - \lambda}/n!$} the probability mass function of the Poisson distribution and $p(x_i|\theta)$ is the likelihood that an event $i$ with reconstruction-level kinematics $x_i$ is described by a certain value of $\theta$ (henceforth simply called the \textit{likelihood}).

Another useful quantity is the score $t(x|\theta)$, which is the first order term of the log-likelihood Taylor expansion. Evaluated at a reference point $\theta_\text{ref}$, it is a sufficient statistic of the likelihood, i.e. a statistically optimal observable around that reference point, and is defined as:
\begin{equation}
    t(x|\theta)_{\theta_{\text{ref}}} = \nabla_\theta \log p(x|\theta)|_{\theta_{\text{ref}}}
\end{equation}

Particle physicists use simulations to generate "synthetic" observations according to a given choice of $\theta$ and compare them to data in order to perform simulation-based inference. Since the MC simulation chain factorizes into the hard-scatter interaction, parton shower and hadronization, and detector simulation, the likelihood can be written as follows

\begin{align}
    p(x|\theta) = \int dz_d \int dz_s \int dz_p \underbrace{p(x|z_{d})p(z_d|z_s)p(z_s|z_p)p(z_p|\theta)}_{p(x,z|\theta)},
    \label{eq:likelihood_factorization}
\end{align}
where: 
\begin{itemize}
    \item $z_p$, $z_s$ and $z_d$ are the latent variables describing the parton-level event (i.e. the parton-level kinematics), the parton shower and hadronization, and the interactions with the detector and event reconstruction, respectively. These variables cannot be directly measured in experiments.
    \item $p(x,z|\theta)$ is the joint likelihood, jointly conditioned on the reconstruction-level variables $x$ and the latent variables of the event, $z = (z_p, z_s, z_d$).
    \item $p(z_s|z_p)$, $p(z_d|z_s)$ and $p(x|z_{d})$ describe the probabilistic evolution from $z_p \to x$ and $p(z_p|\theta)$ is the parton-level likelihood, given by
          \begin{equation}
              p(z_p|\theta) = \frac{1}{\sigma(\theta)} \frac{d \sigma(\theta)}{dz_p},
          \end{equation}
          where $\sigma(\theta)$ and $\frac{d \sigma(\theta)}{dz_p}$ are the total and the differential cross sections, respectively.
\end{itemize}

Since $\sigma(\theta)$ and $\frac{d \sigma(\theta)}{dz_p}$ can be extracted from event generators for an arbitrary $z$ and $\theta$, both the Poisson term and $p(z_p|\theta)$ can be evaluated. However, the simulation of a single event can easily involve many millions of random latent variables. While $p(x|\theta)$ can be sampled \mbox{(${x}\sim p(x|\theta) $)}, calculating the likelihood (or any function of it, such as the score) would require integrating over all the possible histories leading to that observation ($z \rightarrow x$) \cite{Brehmer:2020cvb,PhysRevLett.121.111801}. In other words, its calculation is intractable. Historically, analyses have circumvented this problem by restricting themselves to histograms of one or two physics-motivated variables, thereby reducing their sensitivity. Alternatively, they have used matrix element calculators to build optimal observables, which compress the kinematic information in the event into a single number, but at the cost of having to approximate the simulation chain with a suitable transfer function $p(x|z_p)$, or ignoring it altogether. Examples include the CP-mixing angle used by ATLAS and CMS to study the $H\tau\tau$ coupling~\cite{ATLAS-Htautau,CMS-Htautau1}; an optimal observable defined as the ratio of the matrix element interference term to the SM contribution, used by ATLAS to probe the $HVV$ vertex in $H \to 4\ell$~\cite{ATLAS-H4l}; and a similar matrix element likelihood approach (MELA) adopted by CMS for the same process~\cite{CMS-Htautau2}.

ML-based inference methods have been introduced \cite{Cranmer:2015bka} to estimate likelihood ratios using binary classifiers, with the main drawback that they are quite computationally intensive. Refs. \cite{PhysRevLett.121.111801,Brehmer:2018eca, Brehmer:2018hga} introduced the idea of using augmented data extracted from the simulator as a more efficient approach.

\subsection{Aiding the inference with augmented data}\label{sec:mining_gold}

Even though the likelihood ratio cannot be directly evaluated, the joint likelihood ratio can be calculated since the intractable parts corresponding to the showering and detector effects cancel out,

\begin{align}
\begin{split}
    r(x,z|\theta_0,\theta_1) & \equiv \frac{p(x,z|\theta_0) }{p(x,z|\theta_1)}                                                  \\
                             & =\frac{p(x|z_d)p(z_d|z_s)p(z_s|z_p)p(z_p|\theta_0)}{p(x|z_d)p(z_d|z_s)p(z_s|z_p)p(z_p|\theta_1)} \\
                             & = \frac{p(z_p|\theta_0)}{p(z_p|\theta_1)}                                                        \\
                             & = \frac{d\sigma(z_p|\theta_0)} {d\sigma(z_p|\theta_1)}\frac{\sigma(\theta_1)}{\sigma(\theta_0)},
\end{split}
\end{align}
where $d \sigma (z_p|\theta)$ are the parton-level event weights from the event generator. The parameters $\theta_0$ and $\theta_1$ correspond to the BSM and SM hypotheses, respectively.


Analogously, the joint score can be calculated as follows

\begin{align}
\begin{split}
    t(x,z|\theta) & \equiv \nabla_{\theta} \log p(x,z|\theta)                                                                                        \\
                  & = \frac{ p(x|z_d)p(z_d|z_s)p(z_s|z_p)\nabla_{\theta} p(z_p|\theta)}{ p(x|z_d)p(z_d|z_s)p(z_s|z_p)p(z_p|\theta)}                  \\
                  & = \frac{\nabla_{\theta} d \sigma (z_p|\theta)}{d \sigma (z_p|\theta)} - \frac{\nabla_{\theta} \sigma (\theta)}{\sigma (\theta)}.
\end{split}
\end{align}

The key result derived in \cite{Brehmer:2018eca, Brehmer:2018hga} is that the joint likelihood ratio and joint score can be used to define loss functions that, when minimized, converge to the true likelihood ratio or the true score. This minimization procedure can be achieved by NNs trained with backpropagation and gradient descent. Therefore, a NN can be trained with a suitable loss function to regress the likelihood ratio or score.

There are several variations of this idea, differing mainly in the exact formulation of the employed loss function. The simple but effective SALLY method (\textbf{S}core \textbf{A}pproximates \textbf{L}ikelihood \textbf{L}ocall\textbf{Y}) - a detector-level optimal observable, minimizes the Mean Squared Error (MSE) with respect to the joint score to estimate the score
\begin{align}
\begin{split}
    & L_{\text{SALLY}}[\hat{t}(x)] = \\
    & -\frac{1}{N} \sum_{(x_i,z_i) \sim p(x,z|\theta_\text{ref})} \left| t(x, z|\theta_\text{ref}) - \hat{t}(x|\theta_\text{ref}) \right|^2.
\end{split}
\end{align}

Since this technique is trained to regress the score, $t(x|\theta_\text{ref})$ at the reference point $\theta_\text{ref}$, it is only locally optimal. Furthermore, the SALLY output can be treated like any other observable for performing a likelihood fit. Therefore, this inference method is essentially an estimator of an Optimal Observable which instead of approximating the generation process via a transfer function, learns it implicitly.

Another method is ALICE \cite{Stoye:2018ovl} (\textbf{A}pproximate \textbf{L}ikelihood with \textbf{I}mproved \textbf{C}ross-entropy \textbf{E}stimator), a per-event surrogate model of the likelihood ratio. It uses the joint likelihood ratio and takes as inputs theory parameters at two different values, in addition to the observables $x$. Its loss functional is given by

\begin{align}
\begin{split}
     & L_{\text{ALICE}}[\hat{s}(x|\theta_0,\theta_1)] =                                                                  \\
     & = -\frac{1}{N} \sum_{(x_i,z_i,\theta_\text{choice})} \left[ s(x_i, z_i|\theta_0, \theta_1) \log(\hat{s}(x_i)) \right. \\
     & \left. + (1 - s(x_i, z_i|\theta_0, \theta_1)) \log(1 - \hat{s}(x_i)) \right].
    \label{eq:ALICE}
\end{split}    
\end{align}

Events are drawn according to $\theta_\text{choice}=\theta_1,\theta_0$, i.e. sampling from the SM signal or BSM signal datasets (where the chosen BSM dataset is that with $\theta_\text{benchmark}$ closest to the value of $\theta_0$ being sampled).

In the above formula, $s$ is the boundary classification function modelled by a classifier trained to distinguish between two sets of observations sampled from \mbox{$x\sim p(x|\theta_0)$} and \mbox{$x\sim p(x|\theta_1)$}, and which can be turned into an estimator for the likelihood ratio via the likelihood-ratio trick~\cite{Sugiyama_Suzuki_Kanamori_2012}.

Finally, the ALICES (\textbf{A}pproximate \textbf{L}ikelihood with \textbf{I}mproved \textbf{C}ross-entropy \textbf{E}stimator and \textbf{S}core) method is explored. It is an extension of $L_{\text{ALICE}}$ since it uses not only the joint likelihood ratio but also leverages the joint score to guide the estimator. The loss function is therefore extended as:

\begin{align}
\begin{split}
     & L_{\text{ALICES}}[\hat{s}(x|\theta_0,\theta_1)] = L_{\text{ALICE}} -\frac{1}{N} \sum_{(x_i,z_i,\theta_\text{choice})}\\
     &  \alpha (1 - y_i) \left| t(x_i, z_i|\theta_0, \theta_1) - \nabla_\theta \log \left( \frac{1 - \hat{s}(x_i|\theta, \theta_1)}{\hat{s}(x_i|\theta, \theta_1)} \right) \Big|_{\theta_0} \right|^2 ,
    \label{eq:ALICES}
\end{split}    
\end{align}
where $\alpha$ is a hyper-parameter that weights the joint likelihood ratio and joint score terms in the loss function, $y$ is a label corresponding to $y_i$ = 0 for samples drawn from the BSM hypothesis, $(x_i,z_i) \sim p(x,z|\theta_0)$, and $y_i = 1$ for samples drawn from the SM hypothesis, $(x_i,z_i) \sim p(x,z|\theta_1)$. The ALICES method converges to the ALICE technique when $\alpha=0$ \cite{Stoye:2018ovl}. 

The ALICE/ALICES methods are expected to exhibit superior performance to SALLY as they perform an estimation of the likelihood ratio rather than its local approximation. Since they do not rely on the assumption that the parameter $\theta$ is close to the SM and are parametrized on $\theta$, they are optimal over the entire parameter space. Both the SALLY and ALICE(S) methods are robust against reducible and irreducible backgrounds (which have fixed values of the joint quantities) and to unobservable degrees of freedom, such as the longitudinal momentum of the neutrino in the W decay.

These simulation-based inference techniques based on data augmentation are implemented in a Python package called \texttt{MadMiner} \cite{Brehmer:2019xox}, which was used extensively for this work.

\section{Constraining $HWW$ anomalous
  couplings}\label{sec:analysis}

\subsection{Current state-of-the-art}\label{sec:Effective_Field_Theory}


The SM can be regarded as a low-energy effective theory of an underlying and unknown theory, valid up to an energy scale $\Lambda$ (also called the New Physics scale) beyond the reach of current colliders. In this work, such an effective theory is written using the SMEFT approach, which extends the SM Lagrangian with additional operators of mass dimension $d > 4$, constructed from combinations of SM fields invariant under SM $SU(3)_C \times SU(2)_L \times U(1)_Y$ transformations 

\begin{equation}
    \mathcal{L}_\text{SMEFT} = \mathcal{L}_\text{SM} + \sum_{d > 4} \sum_{i} c_i \frac{O^{(d)}_i}{\Lambda^{d-4}},
\end{equation}
where $c_i$ is called the Wilson coefficient, which parametrizes the relative strength of an operator.

This study focuses on dimension-6 operators. The only dimension-5 operator, the Weinberg operator, is related to neutrino masses and violates lepton number conservation, and is therefore not considered. Using the Warsaw basis \cite{Grzadkowski:2010es}, two dimension-6 operators are pertinent for this work: one CP-odd, which changes sign under a CP transformation, and a CP-even whose sign is not altered under the same transformation. These operators are written as follows
\begin{equation}
    \mathcal{\Tilde{O}}_{HW} = H ^\dag H \epsilon_{\mu \nu \rho \sigma} W^{I \rho \sigma} W ^{I \mu \nu} \ \text{(CP-odd)},
\end{equation}\label{eq:CP_odd_operator}
\begin{equation}
    \mathcal{O}_{HW} =  H ^\dag H {W}^I_{\mu \nu} W^{I \mu \nu} \ \text{(CP-even)},
\end{equation} \label{eq:CP_even_operator}
where $H$ is the $SU(2)_L$ Higgs doublet, $W_{\mu \nu}^a$ the field strength tensor and $\epsilon^{\mu\nu\rho\theta}$ is the antisymmetric Levi-Civita tensor. $c_{H\tilde{W}}$ and $c_{HW}$ are the corresponding Wilson coefficients of the two operators and $\Lambda$ is assumed to be $1 \ \si{TeV}$.

The two operators contribute to the SMEFT amplitude via interference terms with the SM, linearly proportional to $c_i/\Lambda^2$, and via quadratic terms, proportional to $c_i^2/\Lambda^4$. It is worth pointing out that only the linear term of the CP-odd operator is CP-odd. Other higher-order operators can also contribute at the order of $1/\Lambda^4$, but are not considered here for simplicity. Since the true UV theory parametrized by SMEFT is unknown, these coefficients (more particulary, the ratio $c_i/\Lambda^2$) are free parameters and need to be constrained experimentally. 

In this work, different ML-based inference techniques are employed to probe the sensitivity to $c_{HW}$ and $c_{H\tilde{W}}$, aiming to improve the state-of-the-art constraints by ATLAS and CMS, summarized in Table \ref{tab:current_constraints}. The CMS results shown correspond to those available in the Warsaw basis. Other analyses exist in different formalisms, such as the recent simulation-based inference study from CMS~\cite{CMS-WH-SBI}. 


\begin{table}[!htb]
    \centering
    \caption{\small Summary of the most stringent to date expected $68\%$ confidence intervals from ATLAS and CMS for the CP-odd $c_{H\tilde{W}}$ and CP-even $c_{HW}$  Wilson coefficients in the Warsaw basis of SMEFT. The results presented consider the linear plus quadratic effects of new physics.}
    \resizebox{0.49\textwidth}{!}{%
        \setlength{\tabcolsep}{3pt}
        \renewcommand{\arraystretch}{1.1} 
        \begin{tabular}{@{}ccc@{}}
            \hline
            Wilson coefficient & ATLAS (68\% CL) & CMS (68\% CL) \\
            \hline
            $c_{H\tilde{W}}$  & $[-0.22, 0.22]$ \cite{Htautau-CP} & $[-1.11, 1.11]$ \cite{CMS-H4l} \\
            $c_{HW}$  & $[-0.0072, 0.0072]$ \cite{ATLAS:2022tnm}     & $[-0.28, 0.39]$ \cite{CMS-H4l} \\
            \hline
        \end{tabular}}
    \label{tab:current_constraints}
\end{table}

\subsection{Overview of the $WH \rightarrow \ell \nu b \bar{b}$ channel}\label{sec:overview}

\begin{figure}[h!]
    \centering
    \includegraphics[width=0.35\textwidth]{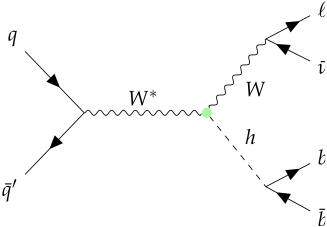}
    \caption{Feynman diagram illustrating the LO $WH$ associated production in the $l \nu b \bar{b}$ final state. The green circle corresponds to the vertex of interest.}
    \label{fig:Wh_production_vlbb}
\end{figure}

This analysis targets the associated $WH$ production channel with the $W$ boson decaying leptonically, \mbox{$ W \rightarrow  l \nu$  ($l = e, \mu$) } and the Higgs boson decaying to a pair of $b$-quarks, $H \rightarrow b \bar{b}$. The final state, illustrated by the leading-order (LO) Feynman diagram in Fig. \ref{fig:Wh_production_vlbb}, is characterized by two jets originating from $b$-quarks, missing transverse energy from the neutrino and either an electron or muon, which enable effective triggering and the removal of most QCD multijet background.

\begin{figure}[h!]
    \centering
    \begin{subfigure}{0.35\textwidth}
        \includegraphics[width=\linewidth]{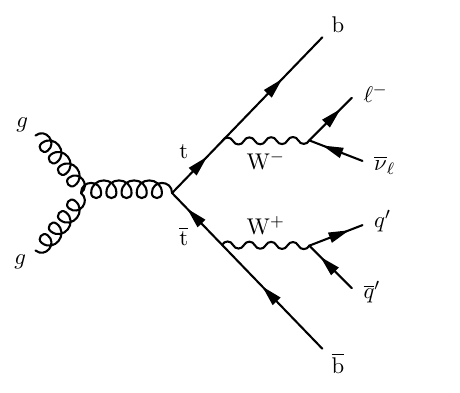}
        \caption{$t\bar{t}$}
    \end{subfigure}
    \begin{subfigure}{0.20\textwidth}
        \includegraphics[width=\linewidth]{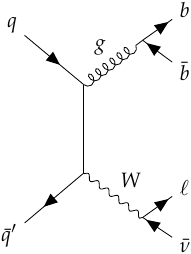}
        \caption{W + $b$-jets}
    \end{subfigure}
    \caption{Leading order Feynman diagram illustrating the most relevant backgrounds.}
    \label{fig:backgrounds}
\end{figure}

The main background processes correspond to top-pair ($t\bar{t}$) production in the semi-leptonic decay channel and associated production of a $W$ boson and $b$-jets. Additional contributions from single top production in the $s$-channel are also considered. The leading-order Feynman diagrams illustrating the most relevant background processes are shown in Fig.~\ref{fig:backgrounds}.

\subsection{Energy-dependent and angular observables}
\begin{figure}[!h]
    \centering
    \includegraphics[width=0.45\textwidth]{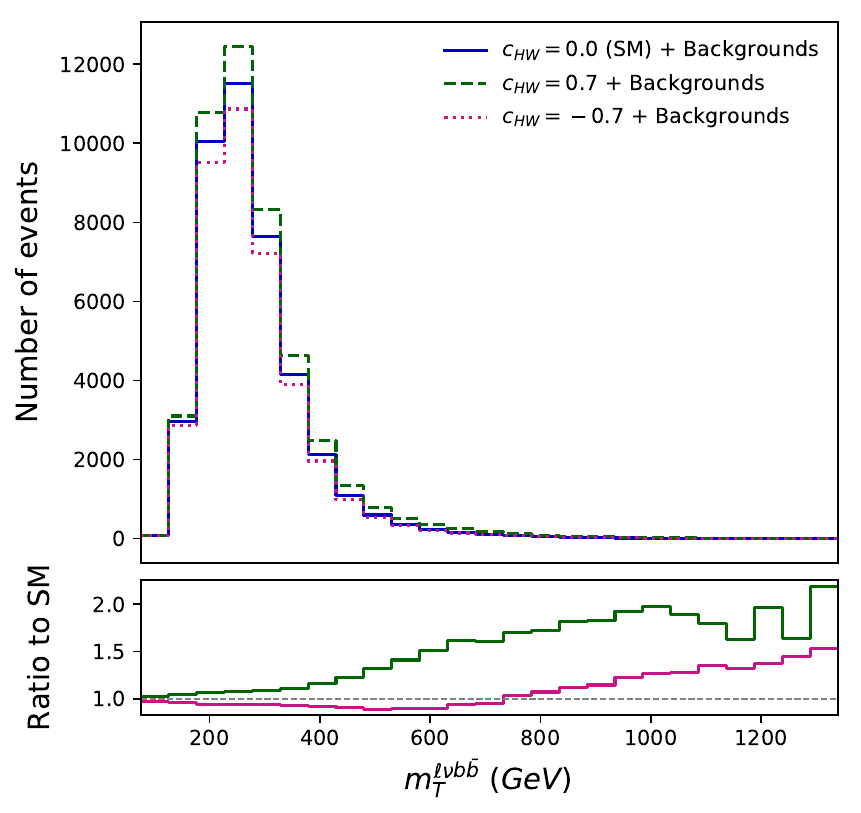}
    \caption{Distributions of the transverse mass of the ${\ell \nu b \bar{b}}$ system for the SM signal plus backgrounds (blue), compared to alternative BSM signal samples plus backgrounds, with $c_{HW} = 0.7$ (dashed green) and $c_{HW} = -0.7$ (dotted pink), obtained using the morphing technique. The lower panel shows the ratio to the SM prediction. The samples and selection cuts are described in Subsection~\ref{subsec:MC}.} \label{fig:mt_tot_cp_even}
\end{figure}

The BSM quadratic components in the amplitude typically lead to more pronounced modifications in the distributions of observables that depend on the partonic centre-of-mass energy. This effect is evident in the distributions of the total transverse mass of the event, $m_T^{\ell \nu b \bar{b}}$, and the transverse momentum of the $W$ boson, $p_T^W$ \cite{sally_cpv_wh, Brehmer:2019gmn}. As illustrated in Figure \ref{fig:mt_tot_cp_even} for $m_T^{\ell \nu b \bar{b}}$, there is an increase in the signal-to-background ratio in the high-energy region compared to the SM prediction, indicating that these observables are sensitive to $c_{HW} \neq 0$. The same is also true for $c_{H\tilde{W}} \neq 0$.

\begin{figure}[!h]
    \centering
    \includegraphics[width=0.45\textwidth]{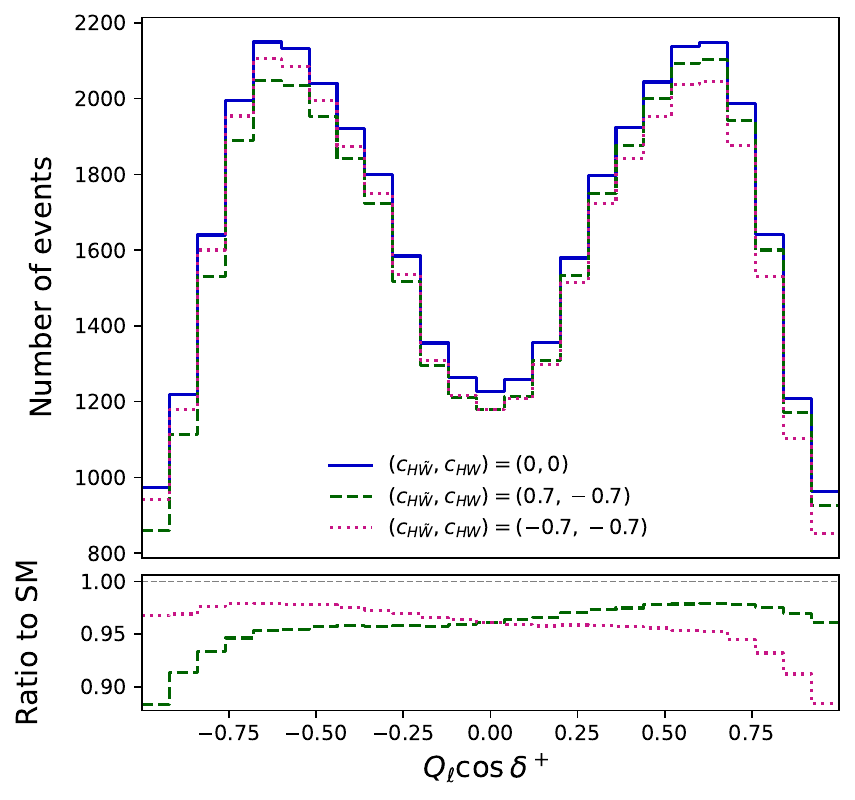}
    \caption{Distributions of the angular variable $Q_\ell \cos \delta^+$ for the SM signal plus backgrounds (blue), compared to alternative BSM signal samples plus backgrounds, with $c_{H\tilde{W}},c_{HW} =(0.7,-0.7)$ (dashed green) and $c_{H\tilde{W}},c_{HW} =(-0.7,-0.7)$ (dotted pink), obtained using the morphing technique. The lower panel shows the ratio to the SM prediction. The samples and selection cuts are described in Subsection~\ref{subsec:MC}.}
    \label{fig:angular_observables}
\end{figure}

However, since these modifications arise primarily from the quadratic term in the amplitude, there is limited sensitivity to the sign of the Wilson coefficients, especially for the CP-odd operator. Therefore, two angular observables~\cite{Godbole:2014cfa,sally_cpv_wh}, particularly sensitive to the CP-odd interference component, were also considered
\begin{equation}
    Q_\ell \cos \delta^+ = Q_\ell \times \frac{\mathbf{p}_\ell^{(W)} \cdot (\mathbf{p}_H \times \mathbf{p}_W)}{|\mathbf{p}_\ell^{(W)}||\mathbf{p}_H \times \mathbf{p}_W|},
\end{equation}
\begin{equation}
    Q_\ell \cos \delta^- =  Q_\ell \times \frac{\mathbf{p}_W \cdot (\mathbf{p}_\ell^{(-)} \times \mathbf{p}_\nu^{(-)})}{|\mathbf{p}_W||\mathbf{p}_\ell^{(-)} \times \mathbf{p}_\nu^{(-)}|},
\end{equation}

where $\mathbf{p}_W$ and $\mathbf{p}_H$ denote the momenta of the $W$ and Higgs bosons, respectively, $\mathbf{p}_\ell^{(W)}$ the momentum of the charged lepton in the $W$ boson rest frame, and $\mathbf{p}_\ell^{(-)}$ and $ \mathbf{p}_\nu^{(-)}$ represent the momenta of the charged lepton and neutrino in the Higgs boson rest frame (with $\mathbf{p}_H \rightarrow - \mathbf{p}_H $). The longitudinal momentum of the neutrino, $p_z^{\nu}$ is reconstructed as described in Ref.~\cite{sally_cpv_wh}. 
The distributions of $Q_\ell \cos \delta^+$ are shown in Figure \ref{fig:angular_observables} for SM and BSM signals plus backgrounds. It is clear that while the distribution remains symmetric around zero for the SM Higgs hypothesis, an asymmetry is introduced for values of $c_{H\tilde{W}}\neq0$, different for a positive or negative value. On the other hand, a positive (negative) value for the CP-even coefficient results in an increase (decrease) of the overall cross-section but no change to the asymmetry of the distribution.

\subsection{EFT scenarios studied}

In this work, the sensitivity of the ALICES, ALICE, and SALLY methods to the CP-odd ($\mathcal{\Tilde{O}}_{HW}$) and CP-even ($\mathcal{O}_{HW} $) dimension-$6$ operators was examined across three different scenarios, in order to evaluate how these inference techniques perform under distinct levels of complexity and sensitivity to BSM couplings. 

\vspace{0.1cm}
\noindent \textbf{Validation at parton level:} 
These methods were first validated in a setup where the likelihood ratio and score can be precisely calculated. By omitting the effects of parton showers, parton distribution functions, and detector responses, all latent variables directly correspond to the observables (i.e., $x=z$). As a result, the likelihood ratio $r(x)$ becomes equivalent to the joint likelihood ratio $r(x,z)$, and the score $t(x)$ matches the joint score $t(x,z)$. Therefore, it is possible to directly compare the neural network predictions with the ground truth. In addition, for the neural network to effectively learn the matrix element, all initial and final state particles need to be fixed, so a simplified parton-level process was considered

\begin{equation}
    u \bar{d} \rightarrow W^+ H \rightarrow \mu^+ \nu_{\mu} b \bar{b}.
\end{equation}
In this scenario, only the $\mathcal{\Tilde{O}}_{HW}$ operator was considered to assess the closure of the methods. \\

\vspace{0.1cm}
\noindent \textbf{Simultaneous 2D fits:} Following the parton-level validation, a more realistic phenomenological analysis is performed by incorporating parton shower and detector effects. 
Specifically, the aforementioned simulation-based inference methods were trained with CP-odd and CP-even coefficients as inputs, in order to investigate their sensivitiy to both operators simultaneously. Two kinematic regimes were considered:  
\begin{enumerate}
    \item \textbf{Inclusive $p_T^W$} — using the full simulated event sample without additional kinematic cuts.  
    \item \textbf{High-$p_T^W$} — selecting events with $p_T^{W} > 150~\text{GeV}$, to probe the enhanced sensitivity to BSM couplings in the high $p_{T}$ regime and to benefit from a higher signal-to-background ratio in this phase-space region.
\end{enumerate}

For comparison, a set of baseline results were established by performing histogram-based fits using energy-dependent and angular observables, namely: a 1D fit to $m_T^{\ell \nu b \bar{b}}$, a 1D fit to $Q_\ell \cos \delta^+$ and a 2D fit to $Q_{\ell}\cos \delta^+ \otimes~p_T^W$. 

All three scenarios were conducted using \texttt{MadMiner v0.9.3} \cite{Brehmer:2019xox}, with distinct simulation setups for each, as detailed in the next Sections.

\subsection{Monte Carlo samples and event selection}\label{subsec:MC}

The signal and background samples are generated following Ref. \cite{sally_cpv_wh}. Signal samples were simulated at $13 \ \si{TeV}$ with \texttt{MadGraph\_aMC@NLO} \cite{Alwall:2014hca} at leading order\footnote{Next-to-leading order corrections for CP-odd SMEFT operators in VH production have been studied in Ref.~\cite{Rossia:2024rfo}, but these are not widely available in event generation tools yet, and are therefore not included in this work.} (LO) in QCD, employing the PDF4LHC15 PDF set. Modifications to the interaction vertex were incorporated using the \texttt{SMEFTsim3} \cite{Brivio2020-dr} UFO model, with the new physics scale set to $\Lambda = 1 \ \si{TeV}$ and assuming $U(3)^5$ flavor symmetry.

Signal events were generated at the SM point $c_{H\tilde{W}}, c_{HW} = (0,0)$ and reweighted to benchmark BSM points, optimized with the \texttt{MadMiner} morphing functionality. Table \ref{tab:benchmarks} summarizes the chosen benchmark points (morphing basis) for all scenarios considered. The Wilson coefficients were limited to a range of $|c_{H\tilde{W}}| < 1.2$ and $|c_{HW}| < 1.0$, slightly less stringent than the current experimental constraints.

To mitigate the large statistical fluctuations that can arise from reweighting events to distant points in parameter space, one-fifth of the signal samples were generated directly at the BSM points \cite{Brehmer:2019xox}.

The background events were generated at LO in QCD using the default \texttt{MadGraph\_aMC@NLO} SM UFO model. No reweighting or morphing was applied to these samples since the CP-odd and CP-even operators do not affect the background processes. The total number of events generated for each scenario is detailed in Table \ref{tab:number_of_generated_events}. 

A set of generator-level selection criteria, originally detailed in Refs. \cite{Brehmer:2019gmn,sally_cpv_wh} was employed, defining the \emph{inclusive } sample. Additionally, \emph{high-$p_{T}^W$} samples were generated with an extra generator-level cut of $p_{T}^{W} > 125$~GeV. This requirement corresponds to cumulative efficiencies of $8.93\%$ for the SM signal, and $0.07\%$, $0.04\%$, and $1.06\%$ for the $t\bar{t}$, W+(b-)jets, and single top backgrounds, respectively. 
Efficiencies for the other generator-level requirements are reported in Ref.~\cite{sally_cpv_wh}.  

Parton showering and hadronization were simulated using \texttt{Pythia8} \cite{SJOSTRAND2015159} with the A14 tune and MLM merging technique, with a merging scale of $Q_{\text{cut}} = 20 \ \si{GeV}$. Detector effects and object reconstruction were simulated using \texttt{Delphes} \cite{deFavereau:2013fsa}, with the ATLAS default card. For both the inclusive and high-$p_T^W$ sample, events with at least two reconstructed anti-$k_t$ jets with radius parameter of 0.6 and $p_{T}>20$~GeV are considered, followed by a requirement that the two leading jets are $b$-tagged with a 70\% efficiency working point, retaining $24\%$ of the total S+B events. For the high-$p_T^W$ sample, an additional cut of $p_T^W > 150$~GeV was applied at reconstruction level, after the inclusive selection, resulting in a cumulative retention of $10\%$ of the generated S+B events.

\begin{table}[h!]
    \centering
    \caption{\small Optimized morphing basis points for the validation and 2D fit scenarios.
For the 2D case, the same points were used for both the inclusive and high-$p_T^W$ studies.}
    \renewcommand{\arraystretch}{1.2}
    \setlength{\tabcolsep}{8pt}
    \begin{tabular}{cccc}
        \hline
        & Validation & \multicolumn{2}{c}{2D} \\
        \cline{2-2} \cline{3-4}
        Coefficient & $c_{H\tilde{W}}$ & $c_{H\tilde{W}}$ & $c_{HW}$ \\
        \hline
        Benchmark 1 & 1.15 & -0.90 & 0.42 \\
        Benchmark 2 & -1.04 & -0.23 & 0.97 \\
        Benchmark 3 & -- & -1.12 & -0.76 \\
        Benchmark 4 & -- & 0.72 & -0.87 \\
        Benchmark 5 & -- & 1.15 & 0.63 \\
        \hline
    \end{tabular}
    \label{tab:benchmarks}
\end{table}

\begin{table}[h!]
    \centering
    \caption{\small Number of generated events for the validation and 2D fit scenarios. 
For the 2D case, the same event samples were used for both the inclusive and high-$p_T^W$ studies.}
    \renewcommand{\arraystretch}{1.1}
    \setlength{\tabcolsep}{13pt}
    \begin{tabular}{lcc}
        \hline
        & Validation & 2D \\
        \hline
        SM Signal     & $1.0 \times 10^6$ & $8.0 \times 10^6$ \\
        Backgrounds   & --                & $12.0 \times 10^6$ \\
        BSM Signal    & $0.2 \times 10^6$ & $8.0 \times 10^6$ \\
        Total         & $1.2 \times 10^6$ & $28.0 \times 10^6$ \\
        \hline
    \end{tabular}
    \label{tab:number_of_generated_events}
\end{table}

The generated cross-sections at different selection stages, namely before any cuts, after generator-level cuts, and after reconstruction-level cuts, together with the corresponding expected event yields for a luminosity of $300 \ \si{fb^{-1}}$, are presented in Tables~\ref{tab:xs_SB_inclusive} and~\ref{tab:xs_SB_highpt} for the $WH$ signal and the main background processes, for the inclusive and high-$p_T^W$ samples, respectively.

\begin{table}[!htb]
    \centering
    \caption{\small Cross-sections at different selection stages and final expected event yields for a luminosity of $300 \ \si{fb^{-1}}$, for the inclusive $p_T^W$ sample.}
    \resizebox{\columnwidth}{!}{%
        \renewcommand{\arraystretch}{1.4}
        \begin{tabular}{@{}lccccc@{}}
            \hline
            Stage & Signal (SM) & $t\bar{t}$ & W+jets & Single top & Total \\ \hline
            No cuts [$\si{fb}$] & $214.00$ & $69310.00$ & $52340.00$ & $666.00$ & -- \\
            After gen-level cuts [$\si{fb}$] & $73.16$ & $194.24$ & $239.50$ & $75.74$ & -- \\
            After reco-level cuts [$\si{fb}$] & $18.29$ & $36.97$ & $63.04$ & $19.54$ & -- \\
            Final events @ $300 \ \si{fb^{-1}}$ & $5485.83$ & $11092.41$ & $18911.46$ & $5861.14$ & $41350.84$ \\ \hline
        \end{tabular}
    }
    \label{tab:xs_SB_inclusive}
\end{table}

\begin{table}[!htb]
    \centering
    \caption{\small Cross-sections at different selection stages and final expected event yields for a luminosity of $300 \ \si{fb^{-1}}$, for the high-$p_T^W$ sample.}
    \resizebox{\columnwidth}{!}{%
        \renewcommand{\arraystretch}{1.4}
        \begin{tabular}{@{}lccccc@{}}
            \hline
            Stage & Signal (SM) & $t\bar{t}$ & W+jets & Single top & Total \\ \hline
            No cuts [$\si{fb}$] & $214.00$ & $69310.00$ & $52340.00$ & $666.00$ & -- \\
            After gen-level cuts [$\si{fb}$] & $19.07$ & $49.57$ & $22.15$ & $7.06$ & -- \\
            After reco-level cuts [$\si{fb}$] & $3.11$ & $1.98$ & $3.31$ & $0.30$ & -- \\
            Final events @ $300 \ \si{fb^{-1}}$ & $934.30$ & $593.92$ & $993.98$ & $91.31$ & $2613.51$ \\ \hline
        \end{tabular}
    }
    \label{tab:xs_SB_highpt}
\end{table}

\subsection{Sample unweighting and augmentation}\label{sec:augmentation}
The loss functions described in Section~\ref{sec:mining_gold} are not well-suited for weighted data. Therefore, an unweighted training sample is created by drawing $N$ events $(x_i, z_i)$ from the original \texttt{MadGraph} samples with probabilities given by their corresponding weights. This is done via inverse transform sampling, where uniformly distributed random variables are converted into samples following $p(x_i, z_i |\theta)$ using the inverse of the cumulative distribution function (CDF). The unweighted sample is then augmented by calculating the joint score and/or the joint likelihood ratio for each event.

For the SALLY method, the sampling process is straightforward: events are sampled at the SM point, and the joint score is evaluated only at this point. In contrast, the ALICE(S) method is more complex, as it involves computing the joint likelihood ratio for multiple pairs of $\theta = (\theta_0, \theta_1)$ values, with $\theta_0$ sampled from either a Gaussian or a flat distribution.

To reduce statistical fluctuations from large event weights, only events generated at the nearest benchmark point were used in the sampling.  The sampling setup for each scenario is summarized in Table \ref{tab:sampling_setup}.

\begin{table}[ht]
    \centering
    \caption{\small Sampling setup for the validation and 2D fit scenarios. 
For the 2D case, the same setup was used for both the inclusive and high-$p_T^W$ studies.}
    \resizebox{\columnwidth}{!}{%
    \begin{tabular}{@{}ccc@{}}
        \hline
        & Validation & 2D \\ \hline
        $n_{\text{samples}}$ & $1.0 \times 10^6$ &  $11.5 \times 10^6$ \\
        Distribution & \begin{tabular}[c]{@{}c@{}}Gaussian ($\mu=0$, $\sigma=0.4$)\\ or Uniform ($[-1.2, 1.2]$)\end{tabular} & \begin{tabular}[c]{@{}c@{}}Gaussian ($\mu=0$, $\sigma=0.4$)\\ or Gaussian ($\mu=0$, $\sigma=0.3$)\end{tabular} \\
        $n_{\theta_0}$ & 1000 or 10000 & 10000 per $\theta$ \\ \hline
    \end{tabular}%
    }
    \label{tab:sampling_setup}
\end{table}

\subsection{Training settings}\label{sec:training}

In this study, the ALICES, ALICE, and SALLY methods were trained for each of the three scenarios using the unweighted training data. The training dataset was defined as 80\% of the total generated samples. This dataset was then further split into 75\% for training and 25\% for validation.

Two sets of variables were used as inputs to the neural networks.  The first set includes kinematic information on the $b$-jets, the charged lepton and missing transverse energy (same variables included in Ref.~\cite{sally_cpv_wh}). The second set is extended to include also the aformentioned angular observables, $Q_\ell \cos \delta^+$ and $Q_\ell \cos \delta^-$, as well as the longitudinal momentum of the neutrino, $p_z^{\nu}$. 

The dense neural network employed in the 2D simultaneous (validation) fit setup consisted of two (one) hidden layers with 100 nodes each (50 nodes) and with Tanh (ReLU) activation functions. The training was performed using \textsc{PyTorch}~\cite{DBLP:journals/corr/abs-1912-01703}, with a batch size of 128, using the AMSGrad~\cite{amsgrad} optimizer with an initial learning rate of 0.001, later reduced to 0.0001. Input variables were standardized to zero mean and unit variance. Early stopping was employed, based on the validation dataset.

To improve the robustness of the result, an ensemble of five neural networks was trained for each setup, each using a separate unweighted dataset obtained through independent sampling from the same initial weighted dataset. In the case of ALICES, the additional hyperparameter entering the loss function $\alpha$ had to be defined and was set to 5 for the validation and high-$p_T^W$ scenarios and 10 for inclusive $p_T^W$. These values were found to reduce the variance of the individual estimators entering the ensemble, as later discussed.

\subsection{Limit setting}

To extract the 68\% and 95\% confidence levels (CL) of the $c_{H\tilde{W}}$ and $c_{HW}$ coefficients, the $p$-values associated with the negative-log-likelihood-ratio test statistic were calculated in the asymptotic limit, across the $\theta$ parameter space, and using an `Asimov dataset' \cite{Cowan2010-ox}.

For ALICE(S), calculating the $p$-values is straightforward, as it involves evaluating the NNs at various parameter points to obtain the likelihood ratio. For the SALLY technique, inference is performed similarly to histograms of summary statistics. The $c_{H\tilde{W}}$ coefficient was scanned over a parameter range of $[-1.2,1.2]$ and $c_{HW}$ over $[-1.0,1.0]$, using 35 points in each direction. The Asimov datasets (the template histograms) were built with 50\% (100\%) of the dataset. 

The template histograms for $m_T^{\ell \nu b \bar{b}}$, $Q_\ell \cos \delta^+$, and $Q_{\ell}\cos \delta^+ \otimes p_T^W$ were built following the binning of Ref.~\cite{sally_cpv_wh}. For SALLY, $8 \times 8$ bins were used in the two-dimensional parameter space. A luminosity of $300 \ \si{fb}^{-1}$ was assumed for the analysis.

\section{Results}\label{sec:results}
\subsection{Validation at parton-level}

\begin{figure*}[t]
    \centering
    \includegraphics[width=.333\linewidth]{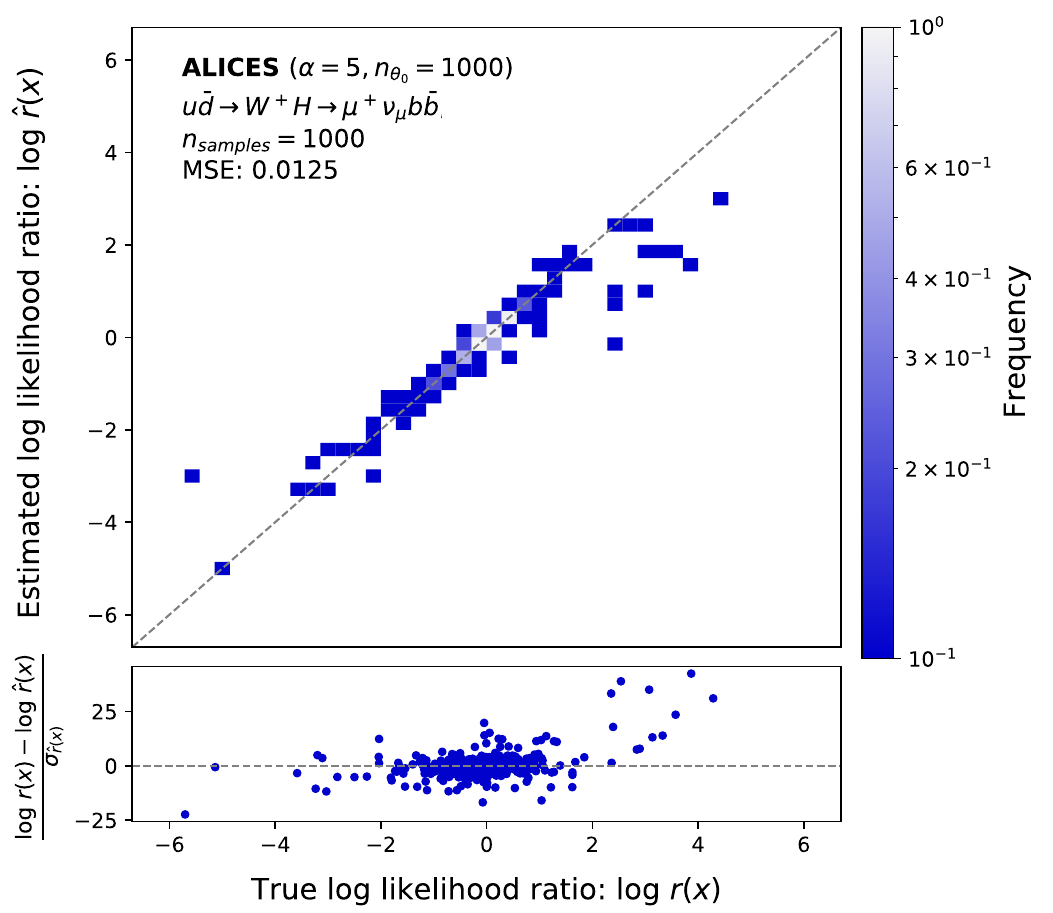}\hfill
    \includegraphics[width=.333\linewidth]{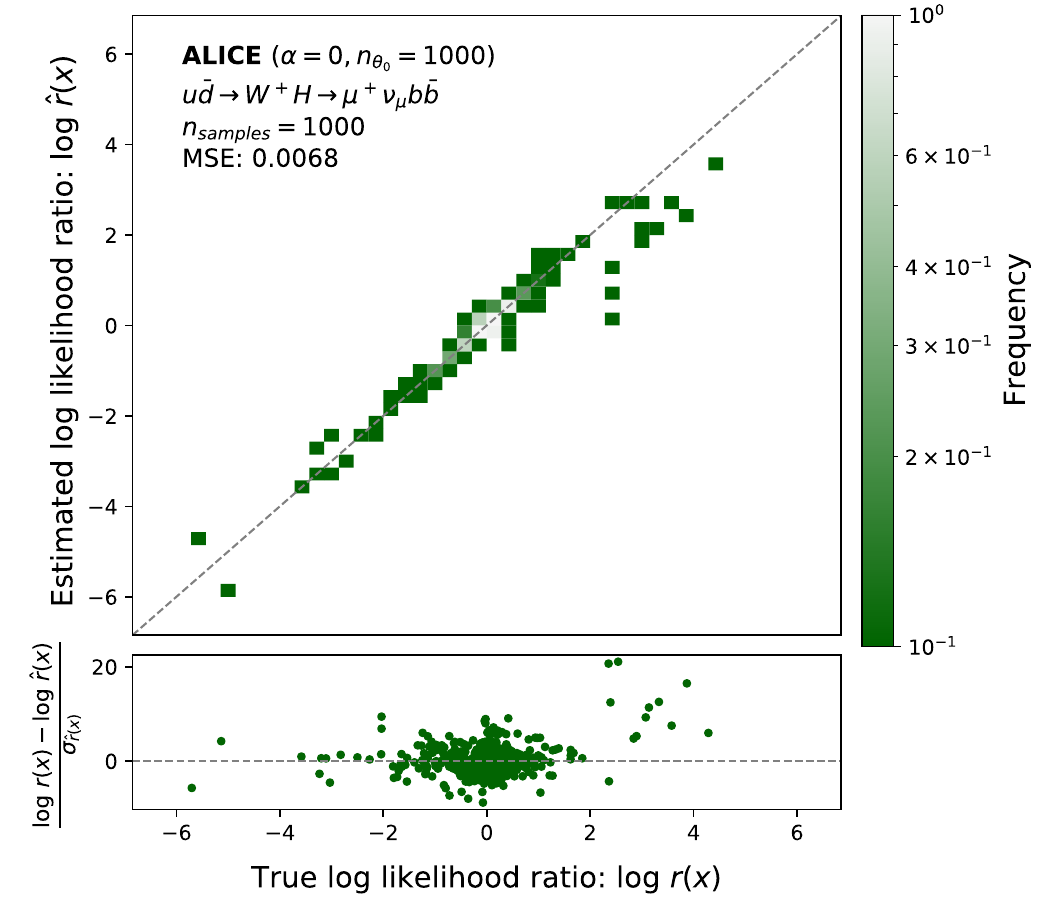}\hfill
    \includegraphics[width=.333\linewidth]{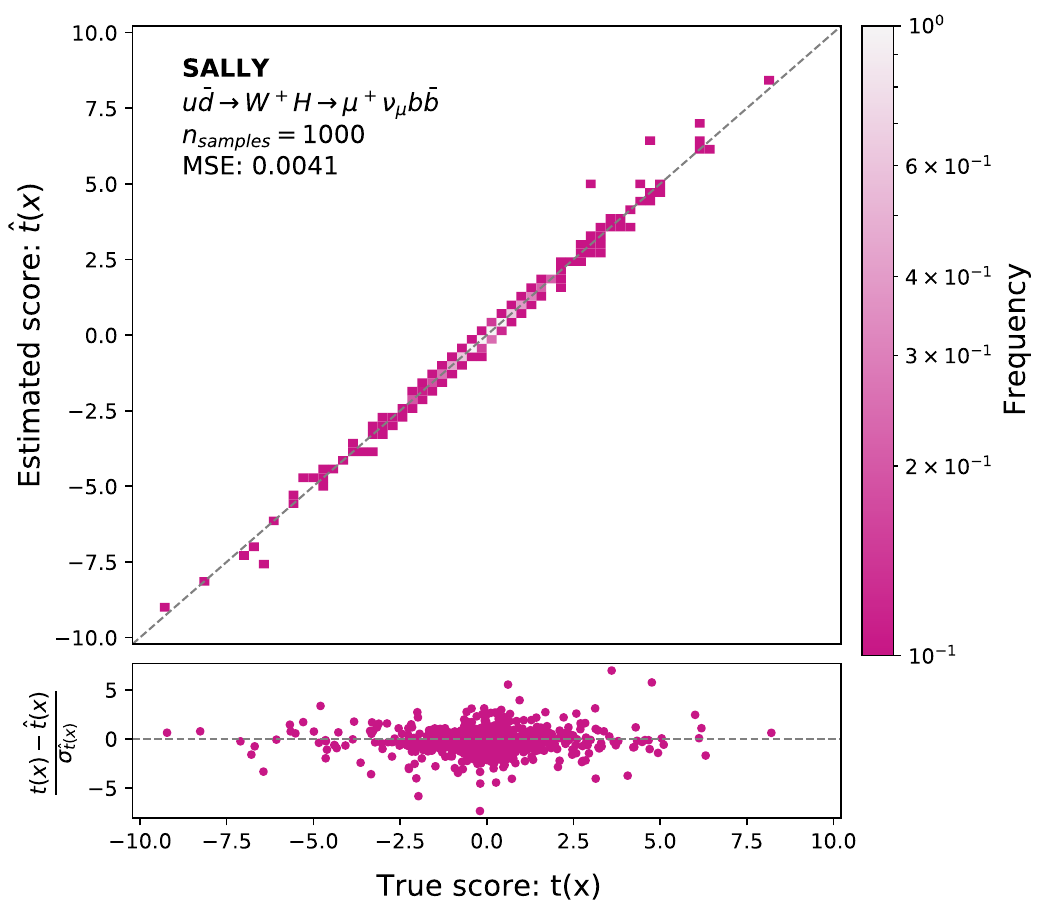}
    \caption{Estimated vs true log-likelihood ratio/score for the ALICES (left), ALICE (middle), and SALLY (right) techniques.}
    \label{fig:estimated_vs_true}
\end{figure*}

To validate the three methods, estimated versus true likelihood ratio/score curves were plotted, and the Mean Squared Error (MSE) was computed to quantify the degree of agreement between the two quantities. The values obtained were 0.0125 for ALICES, 0.0068 for ALICE and 0.0041 for SALLY.  As illustrated in Figure \ref{fig:estimated_vs_true}, the estimated quantities from each method closely align with the true values, confirming that these inference techniques yield reliable results, at least within the truth-level scenario.

In addition, the resulting MSE values for the ALICE and ALICES methods are consistently higher when trained with a Uniform distribution, since it leads to a sparser sampling across the $\theta$ parameter space. Therefore, a Gaussian distribution, which is more sample efficient, was chosen for all subsequent studies.

Regarding the individual negative-log-likelihood scans, the ALICES and ALICE estimator exhibit significant variance among their predictions for the maximum likelihood estimate (MLE). As prescribed in Ref.~\cite{Brehmer:2018eca}, constructing an ensemble proved essential to extract the correct minimum when using the ALICE(S) techniques. For example, the standard deviation in the central values of 5 ALICE(S) estimators was found to be 0.6 (0.2) compared to 0.001 from SALLY.

Upon some investigation of the potential sources of the ALICE(S) variance, it was found that a portion of it can be attributed to the different $\theta_0$ populations within each dataset used for training. This was observed to be slightly mitigated by increasing the number of $\theta_0$ drawn from the Gaussian distribution, specifically from 1000 to 10000. 
A non-negligible variance remained, which can in part be attributed to the increased complexity of the ALICE(S) loss functions compared to SALLY: by learning the $\theta$ dependence, these models become more susceptible to outliers and fluctuations. The value of the hyperparameter $\alpha$ was also found to have some impact on the resulting variance. Finally, an additional source of variance is the contribution of events from a phase-space regions with low signal to background (S/B) ratio. This is further discussed in the next Section. 

\subsection{2D simultaneous fit}

SALLY, ALICE and ALICES were trained with both the kinematic-only and extended sets of input features. It was observed that including the angular observables and the longitudinal momentum of the neutrino helped the estimation tasks. It was also observed that the ALICE method failed to learn the likelihood function, exhibiting a double minima behavior with very large variance between the estimators\footnote{This behavior was observed even in the ideal scenario of signal-only inference.}. This highlights the importance of including the joint score information, as is done in ALICES. 

\begin{figure*}[t]
    \centering
    \begin{subfigure}{0.45\textwidth}
        \includegraphics[width=\textwidth]{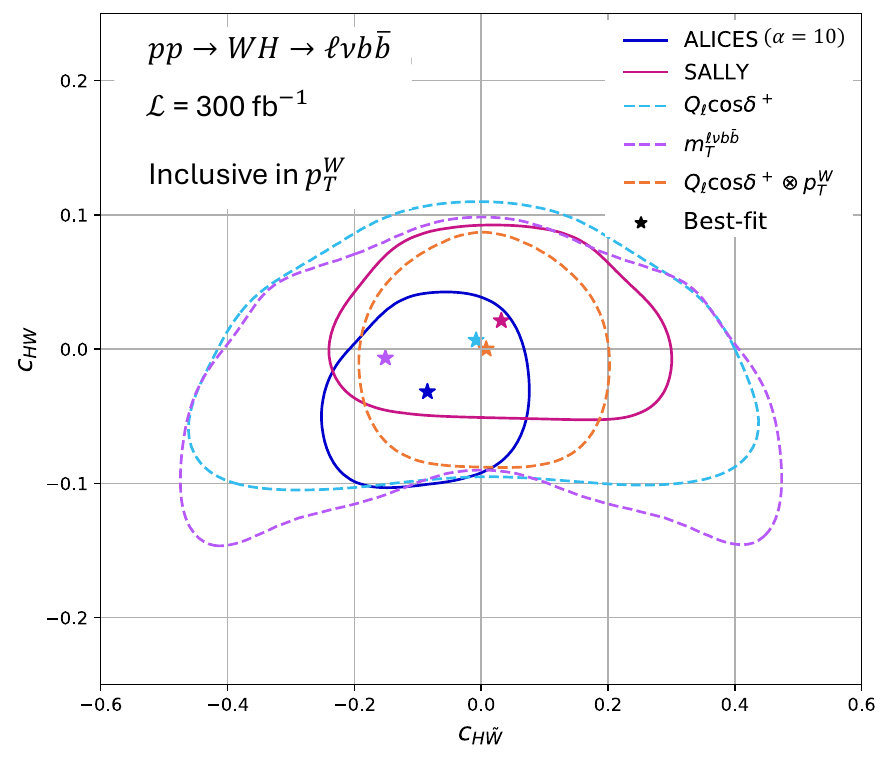}
        \caption{}
    \end{subfigure}
    \begin{subfigure}{0.45\textwidth}
        \includegraphics[width=\textwidth]{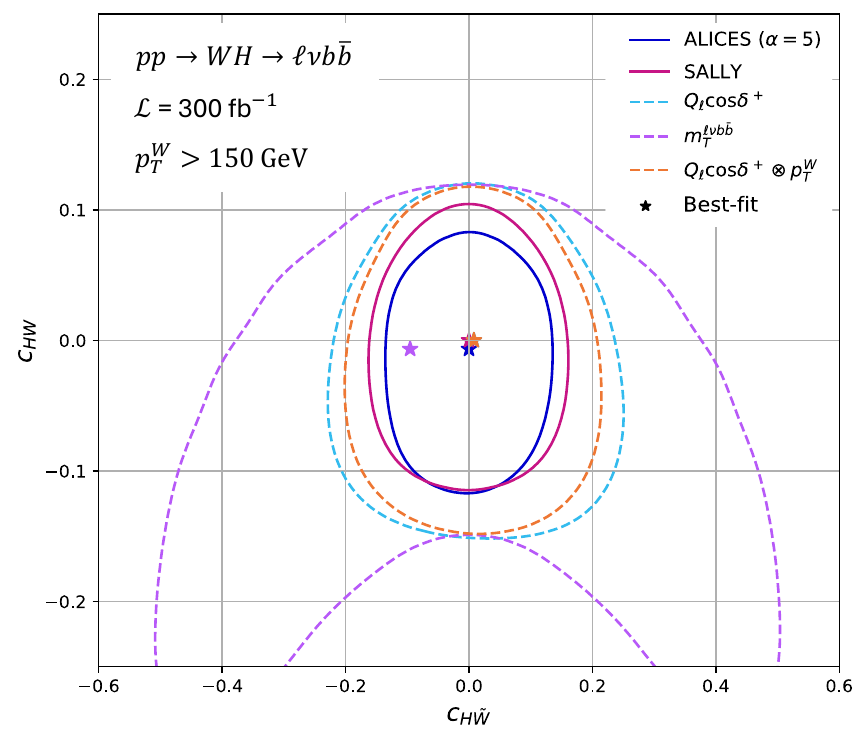}
        \caption{}
    \end{subfigure}
    \caption{Likelihood ratio contours at 95\% CL in a simultaneous fit of the Wilson coefficients $c_{H\tilde{W}}$ and $c_{HW}$.  The scans compare the ALICES (blue) and SALLY (pink) methods, as well as the histogram-based results with $Q_{\ell} \cos \delta^+$, $m_T^{\ell \nu b \bar{b}}$ and $Q_{\ell} \cos \delta^+ \otimes p_T^W$. The best-fit points are marked with a star. The models are trained and evaluated in (a) an inclusive phase-space and in (b) a subset of events where $p_{T}^W>150$~GeV.}
    \label{fig:results_2d}
\end{figure*}

The results are shown in Figure~\ref{fig:results_2d}. Overall, SALLY and ALICES yielded tighter limits than those obtained with the best-performing one-dimensional summary statistic 
($Q_{\ell}\cos \delta^+$ for $c_{H\tilde{W}}$ and $m_T^{\ell \nu b \bar{b}}$ for $c_{HW}$). 
However, the constraints from SALLY were nearly identical to those obtained using the two-dimensional histogram of 
$Q_{\ell}\cos \delta^+ \otimes p_T^W$. 
This is illustrated in Figure~\ref{fig:results_2d}(a), which shows the 95\% confidence level (CL) contours in the $c_{H \tilde {W}}$–$c_{HW}$ plane. 
Nevertheless, SALLY remains particularly promising compared to this 2D histogram, as it can be straightforwardly extended to probe more couplings simultaneously, i.e. without the need for defining new observables.

\begin{figure*}[t]
    \centering
    \begin{subfigure}{0.45\textwidth}
        \includegraphics[width=\textwidth]{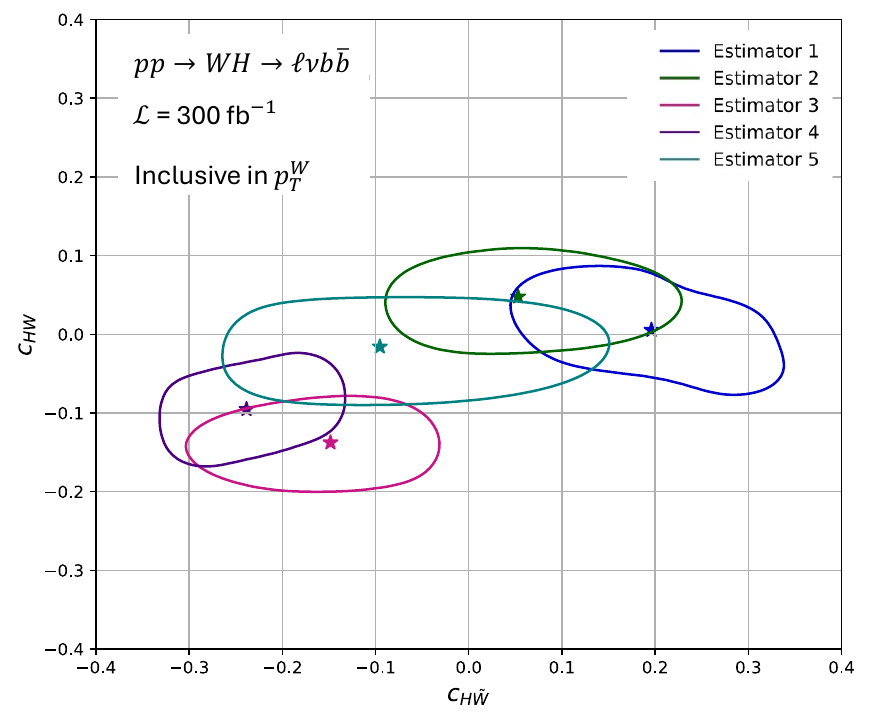}
        \caption{}
    \end{subfigure}
    \begin{subfigure}{0.45\textwidth}
        \includegraphics[width=\textwidth]{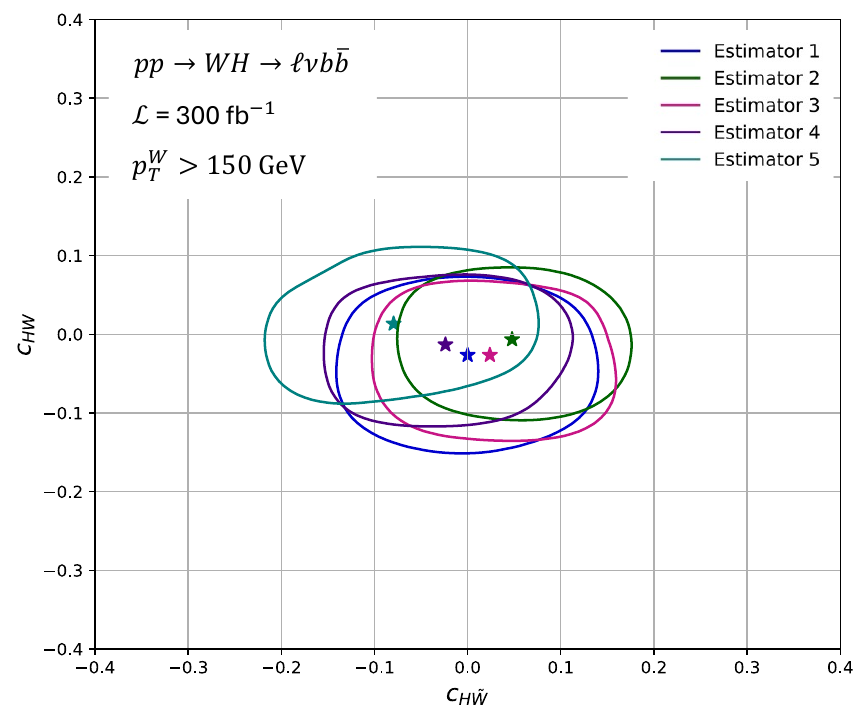}
        \caption{}
    \end{subfigure}
    \caption{Likelihood ratio contours at 95\% CL in a simultaneous fit of the Wilson coefficients $c_{H\tilde{W}}$ and $c_{HW}$.  The scans compare the independent ALICES estimators in (a) an inclusive phase-space and in (b) a subset of events where $p_{T}^W>150$~GeV. The hyperameter $\alpha$ is set to 10 and 5, respectively. The best-fit points are marked with a star.}
    \label{fig:individual_estimators}
\end{figure*}

\begin{figure}[]
    \centering
    \includegraphics[width=0.8\linewidth]{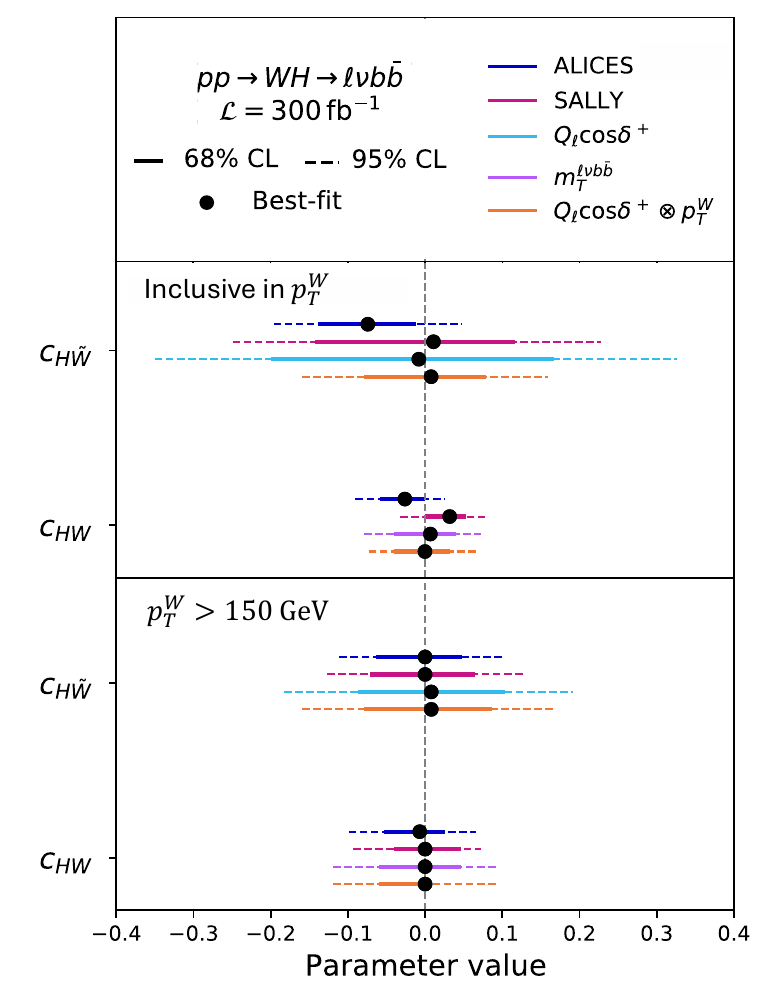}
    \caption{Summary of the 68\% and 98\% CLs obtained for $c_{H \tilde{W}}$ and $c_{HW}$ Wilson coefficients. The results are shown for the fits performed in an inclusive phase-space (top) and in a subset of events where $p_{T}^W>150$~GeV (bottom).}
    \label{fig:summary_2d}
\end{figure}

In order to understand the role of the low S/B regions in the performance of the methods, the training and evaluation were repeated with events in the $p_T^{W}>150$~GeV region. From the results shown in Figure~\ref{fig:results_2d}(b), it is evident that focusing on the phase-space with higher S/B is helpful in reducing the variance of the ML methods and producing the correct minimum. This is especially true for ALICES, as shown in Figure~\ref{fig:individual_estimators}, where the ALICES contours are plotted for each estimator of the ensembles, comparing the inclusive and the high-$p_{T}^W$ regions. In terms of sensitivity to the CP-odd operator, for both ALICES and SALLY focusing on the high-$p_{T}^W$ regions produces tighter constraints. On the other hand, focusing on this region results in a lower sensitivity to the $c_{HW}$ coefficient, for all methods, but especially the $m_T^{\ell \nu b \bar{b}}$ histogram. This behavior is explained by the distinct impact of the CP-even and CP-odd operators as a function of $p_{T}^W$: higher absolute values of $c_{H\tilde{W}}$ lead to increasingly higher S/B as $p_{T}^W$ increases, independently of the sign of the coefficient; on the other hand, the impact of $c_{HW}$ on the $p_{T}^W$ distribution is less pronounced and depends on its sign. A positive CP-even coefficient impacts mostly the signal normalisation, while a negative coefficient can increase (lower) the S/B in the high (low) $p_{T}^W$ region. This is the same behavior observed in the histogram of the (highly correlated) variable $m_T^{\ell \nu b \bar{b}}$ (see Figure~\ref{fig:mt_tot_cp_even}). Finally, it is interesting to observe that the sensitivities of the ML methods to the CP-even coefficient are less susceptible to this narrowing of the phase-space: this is potentially another advantage of further exploring them. The obtained 95\% individual intervals, calculated by fixing one Wilson coefficient to zero while profiling over the other, are shown in Tables~\ref{tab:results_table1} and~\ref{tab:results_table2} for all methods and summarised in Figure~\ref{fig:summary_2d}. We see that in the high-$p_T^W$ region, ALICES (SALLY) produces limits on $c_{H\tilde{W}}$ that are 1.6 (1.3) times better than the 2D summary statistics. For $c_{HW}$, SALLY produces the best limits in the high-$p_T^W$ region, 1.3 times tighter than the 2D histogram (1.2 times for ALICES). For the inclusive scenario, leaving out ALICES given its less robust results, we see that the 2D summary statistic outperforms SALLY by a 1.2 factor for $c_{H\tilde{W}}$, whereas SALLY outperforms the 2D histograms by a factor of 1.3 for the $c_{HW}$ coefficient.

The strongest limits on $c_{H\tilde{W}}$ and $c_{HW}$ reported above can be used to derive 95\% CL lower limits on the scale of new physics, $\Lambda$, corresponding to 38 and 53 TeV, respectively.



\begin{table}[!htb]
    \centering
    \caption{\small Summary of the 95\% CLs for the $c_{H\tilde{W}}$ Wilson coefficient, obtained using the ALICES and SALLY methods, as well as the best limits derived from 1D and 2D summary statistics: $Q_{\ell} \cos \delta^+$ and $Q_{\ell} \cos \delta^+ \otimes p_T^W$. The results are shown for the fits performed in an inclusive phase-space and in a subset of events where $p_T^W > 150 \ \si{GeV}$.}
    \resizebox{0.4\textwidth}{!}{%
        \setlength{\tabcolsep}{8pt}
        \renewcommand{\arraystretch}{1.1} 
        \begin{tabular}{@{}ccc@{}}
            \hline
             & Inclusive in $p_T$ & $p_T^W > 150 \ \si{GeV}$ \\
            \hline
            ALICES  & [-0.18, 0.11] & [-0.11, 0.10] \\
            SALLY   & [-0.19, 0.20] & [-0.13, 0.13] \\
            $Q_{\ell} \cos \delta^+$   & [-0.35, 0.34] & [-0.18, 0.19]\\
            $Q_{\ell} \cos \delta^+ \otimes p_T^W$   & [-0.15, 0.17] & [-0.16, 0.17] \\
            \hline
        \end{tabular}
    }
    \label{tab:results_table1}
\end{table}

\begin{table}[!htb]
    \centering
    \caption{\small Summary of the 95\% CLs for the $c_{HW}$ Wilson coefficient, obtained using the ALICES and SALLY methods, as well as the best limits derived from 1D and 2D summary statistics: $m_T^{\ell \nu b \bar{b}}$ and $Q_{\ell} \cos \delta^+ \otimes p_T^W$. The results are shown for the fits performed in an inclusive phase-space and in a subset of events where $p_T^W > 150 \ \si{GeV}$.} 
    \resizebox{0.4\textwidth}{!}{%
        \setlength{\tabcolsep}{6pt}
        \renewcommand{\arraystretch}{1.1} 
        \begin{tabular}{@{}ccc@{}}
            \hline
             & Inclusive in $p_T$ & $p_T^W > 150 \ \si{GeV}$\\
            \hline
            ALICES  & [0.05, 0.14] & [-0.10, 0.07] \\
            SALLY   & [-0.05, 0.06] & [-0.09, 0.07] \\
            $m_T^{\ell \nu b \bar{b}}$   & [-0.08, 0.07] & [-0.12, 0.09]\\
            $Q_{\ell} \cos \delta^+ \otimes p_T^W$   & [-0.07, 0.07] & [-0.12, 0.09] \\
            \hline
        \end{tabular}
    }
    \label{tab:results_table2}
\end{table}

\subsection{Discussion}

\begin{figure}[h!]
    \centering
    \includegraphics[width=0.8\linewidth]{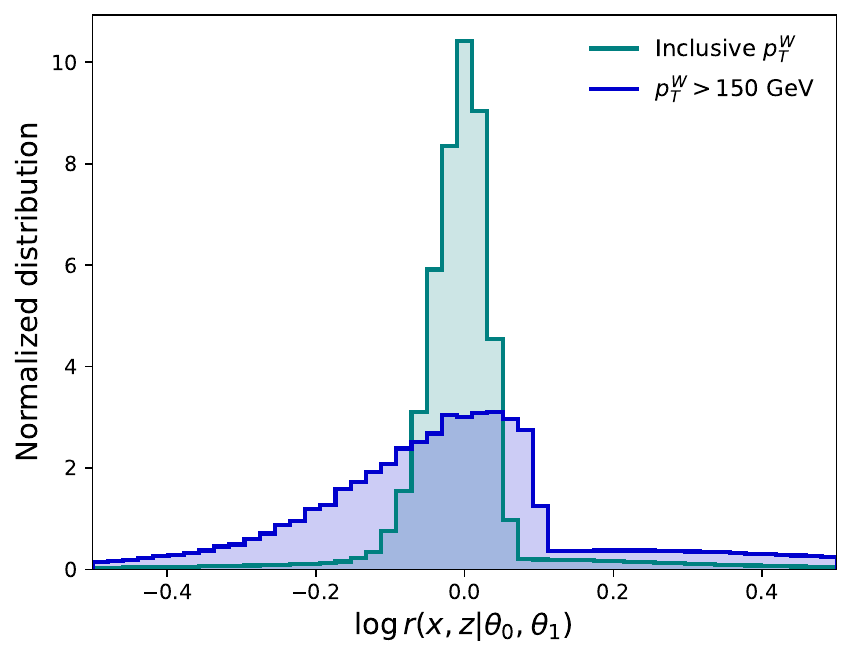}
    \caption{Histograms of the joint log-likelihood ratio for the inclusive (teal) and high-$p_T^W$ (blue) scenarios.}
    \label{fig:joint_llr_hist_cp_odd}
\end{figure}

The ALICES estimators learn the correct marginalization of $r(x,z|\theta_0, \theta_1)$ according to the joint density $p(x,z)$. Therefore, in order to better understand the observed performance, it is useful to plot the joint likelihood ratio for several values of $\theta_0$ drawn from a Gaussian distribution. This is effectively a snapshot of the augmented data the network uses as input. Figure \ref{fig:joint_llr_hist_cp_odd}, shows the histograms of the joint likelihood ratio for both the inclusive and high-$p_T^W$ scenarios.

From Figure \ref{fig:joint_llr_hist_cp_odd}, it is clear that the joint quantity takes extremely small values, particularly for the inclusive $p_T^W$ scenario, confirming how kinematically alike the BSM parameter points are. Therefore, any bounds derived from this quantity will be highly sensitive to minor variations in the network output. One way to tackle this is naturally to reduce the impact of the backgrounds and increase the sensitivity to the BSM couplings, as is done by moving to higher $p_{T}^W$ regions. 

\begin{figure*}[t]
    \centering
    \begin{subfigure}{0.45\textwidth}
        \includegraphics[width=\textwidth]{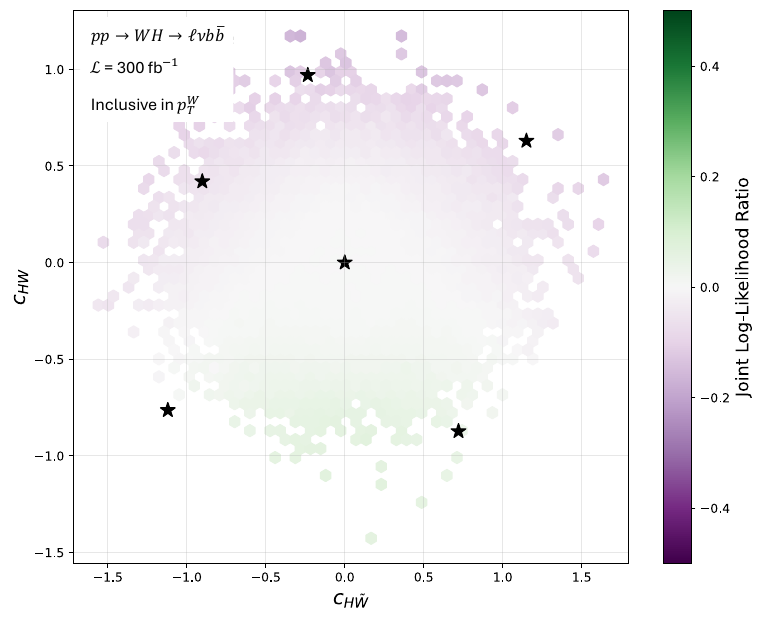}
        \caption{}
    \end{subfigure}
    \begin{subfigure}{0.45\textwidth}
        \includegraphics[width=\textwidth]{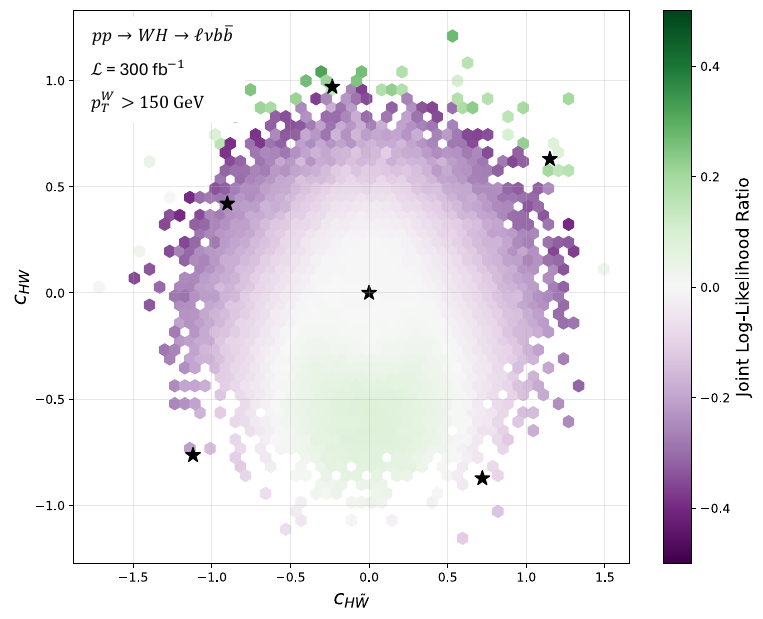}
        \caption{}
    \end{subfigure}
    \caption{Density plots of the joint log-likelihood ratio for the ALICES method in the $(c_{H\tilde{W}}, c_{HW})$ parameter space. The plots are shown for (a) an inclusive phase-space and (b) a subset of events where $p_{T}^W>150$~GeV.}\label{fig:joint_llr_density}
\end{figure*}

The noticeable improvement in performance in the high-$p_T^W$ scenario, compared to the inclusive one, can be understood by examining the 2D density plots of the joint log-likelihood ratio over the $(c_{H\tilde{W}}, c_{HW})$ parameter space, shown in Figure~\ref{fig:joint_llr_density}. For the inclusive case (Figure~\ref{fig:joint_llr_density}a), the joint log-likelihood ratio is nearly flat across the parameter space, making it extremely challenging for the NN to distinguish between different values of the Wilson coefficients. This explains the observed spread in the MLEs and the significant variance among the individual estimators. 

In contrast, for the high-$p_T^W$ region (Figure~\ref{fig:joint_llr_density}b) a clear radial spread is visible: the magnitude of the joint log-likelihood ratio increases with the absolute values of the Wilson coefficients, reflecting the enhanced sensitivity to BSM effects. The two main populations in the plot also become more pronounced: the green region, corresponding to points where the linear term in $c_{HW}$ dominates and, being negative, leads to cross sections smaller than the SM prediction; and the purple region, where the cross section increases relative to the SM. Near the SM point, the joint log-likelihood ratio remains close to zero, as expected from the small impact of the BSM parameters. The enhanced discriminating power across the parameter space in the high-$p_T^W$ case enables the estimator to more reliably distinguish between SM and BSM scenarios, improving the overall sensitivity of the method.

It's noteworthy that the SALLY estimator, on the other hand, learns the score function, which is an optimal observable that encapsulates all the information about the parameters near the SM and, therefore, can effectively capture the differences between SM and BSM points, even for small values of the Wilson coefficients.

From a practical perspective, the computational cost of implementing these ML-based methods can be substantial, particularly for ALICES, where training and evaluation can take several hours, in contrast to the few minutes required for SALLY. In addition, the sampling time for SALLY is also significantly shorter, since it only requires events generated at the SM point. A possible way to improve the sampling efficiency of ALICES would be to calculate a binned cumulative distribution function or to explore alternative sampling strategies. These approaches are left for future work.

Regarding the robustness of the methods of ALICE(S), in addition to ensembling, calibration methods can also be explored, with the general aim of ensuring better estimators of the true likelihood ratios than the raw output of the machine learning models. In particular, the expectation calibration method described in Ref.~\cite{Brehmer:2018eca} is a type of closure calibration to ensure that evaluating the likelihood ratio estimator on a sample of events drawn according to the SM hypothesis returns an expectation value of 1. This was studied but found to have no significant impact on the results. Another calibration method is to ensure e.g. that the classifier $\hat{s}(x|\theta_0,\theta_1)$ has a direct probabilistic interpretation and corresponds to the optimal decision function as expected by the likelihood ratio trick. A challenge with pursuing such a calibration in this case is that it has to be implemented for the various $\theta$ points. This and other possible strategies for improving robustness have been left for future work.

Finally, a possible simplification of the EFT problem would be to factorize out the dependence on the Wilson coefficients and implement one estimator per contribution in the squared matrix element. With this modification, a parametrized classifier is no longer necessary. ALICES would still be a good candidate to derive the resulting estimators, given the documented improvement in computational efficiency when compared to a method like CARL based on a standard cross-entropy estimator~\cite{Stoye:2018ovl}.

Overall, these results demonstrate that such ML-based techniques hold significant promise for application in realistic LHC analyses. However, this study highlights the importance of further investigating ways to improve their robustness, understanding their behaviour across different physics scenarios, and optimising their implementation for both accuracy and computational efficiency.

\section{Conclusions}\label{sec:conclusion}

In this work, the sensitivity to the $c_{H\tilde{W}}$ and $c_{HW}$ coefficients was explored using ML-based inference methods, in the $WH \rightarrow \ell \nu b\bar{b}$ channel ($\ell = e, \mu$).

As the LHC Run 3 advances, the application of ML-based inference methods, such as ALICES and SALLY, shows great promise in probing $HWW$ anomalous couplings with higher sensitivity and precision. In the $WH \rightarrow \ell \nu b\bar{b}$ channel, and focusing on the high-$p_T^W$ region, we observe that ALICES (SALLY) produces limits on $c_{H\tilde{W}}$ that are 1.6 (1.3) times better than the 2D summary statistics. When it comes to the CP-even operator $c_{HW}$, it is SALLY that produces the best limits in this region, 1.3 times tighter than the 2D histogram (1.2 times for ALICES). These techniques offer the potential to improve upon the traditional methods and current results from the ATLAS and CMS collaboration.

The advantages of these techniques come with the trade-off of increased complexity and resource demands. In particular, large amounts of training data are needed to effectively train neural networks, and it can be difficult for them to converge to the true likelihood ratio when BSM signals are similar to the SM ones or in low S/B scenarios. In the particular channel studied in this work, a cut on $p_{T}(W)$ was essential to improve the robustness of the ALICES estimators. 

Despite these complexities, their potential for simulatenously probing multiple SMEFT operators will make them highly valuable tools for Run 3 and beyond. Therefore, this work highlights the importance of addressing the shortcomings of these techniques, such as training stability and computational efficiency, to fully realize their potential.

More broadly, an increase in the sensitivity to the studied Wilson coefficients is directly connected to an increase in the sensitivity to models with extended scalar sectors such as the Two Higgs Doublet Model (2HDM) and the Complex Two Higgs Doublet Model (C2HDM), which can generate non-zero values for $c_{HW}$ \cite{EFTDiagrammaticaCPeven} and $c_{H\tilde{W}}$ \cite{CPoddHWW_BSM}, respectively. Ref. \cite{CPoddHWW_BSM} shows that a value of $c_{H\tilde{W}} \approx 10^{-3}$ can be generated in the C2HDM, which may be attainable at future colliders.

Finally, beyond $c_{HW}$ and $c_{H\tilde{W}}$, the strength of these methods in probing simultaneously a large number of operators will bring sensitivity to a wider range of new physics such as models with scalar leptoquarks or vector-like fermions, which can generate non-zero values for a large range of CP-even \cite{EFTDiagrammaticaCPeven} and CP-odd \cite{EFTDiagrammaticaCPOdd} operators, respectively.

\section*{Acknowledgments}
The authors would like to thank Johann Brehmer and Aishik Ghosh for helpful discussions and feedback on the manuscript. M. S. is supported by Nikhef. I. O. is supported by the Portuguese funding agency FCT (Fundação para a Ciência e a Tecnologia) within projects No. 2023.00042.RESTART and 2023.06052.CEECIND/CP2838/CT0001. R. B. is supported by the Austrian Science fund (FWF) under Grant No. PAT7453824 and received support from FCT within Projects No. SFRH/BD/150792/2020 and 2023.00042.RESTART. The authors would also like to acknowledge the FCT Project No. 2024.00227.CERN. 

\section*{Data Availability Statement}

The code used throughout this analysis is publicly available in Ref.~\cite{silva_hww_code}. The samples are available upon request.

\bibliography{Bibliography_DB}

\end{document}